\documentclass[twocolumn]{aastex62}

\graphicspath{{./}{figures/}}

\usepackage{xcolor}
\def\mearth{{\rm\,M_\oplus}}

\shorttitle{Asteroid belt excitation}
\shortauthors{Deienno et al.}

\begin{document}

\title{The excitation of a primordial cold asteroid belt as an outcome of the planetary instability}

\author{Rogerio Deienno}
\affil{Department of Space Studies, Southwest Research Institute, 1050 Walnut St., Boulder, CO 80302, USA}
\affil{Instituto Nacional de Pesquisas Espaciais, Avenida dos Astronautas 1758, CEP 12227-010 S\~ao Jos\'e dos Campos, SP, Brazil}
\correspondingauthor{Rogerio Deienno}
\email{rdeienno@swri.boulder.edu}

\author{Andr\'e Izidoro}
\affil{Universidade Estadual Paulista, Grupo de Dinamica Orbital e Planetologia, Av. Dr. Ariberto Pereira da Cunha, 333, CEP 12516-410, Guaratinguet\'a, SP, Brazil}

\author{Alessandro Morbidelli}
\affil{Laboratoire Lagrange, UMR7293, Universit\'e C\^ote d'Azur, CNRS, Observatoire de la C\^ote d'Azur, Boulevard de l'Observatoire, 06304 Nice Cedex 4, France}

\author{Rodney S. Gomes}
\affil{Observat\'orio Nacional, Rua General Jos\'e Cristino 77, CEP 20921-400 Rio de Janeiro, RJ, Brazil}

\author{David Nesvorn\'y}
\affil{Department of Space Studies, Southwest Research Institute, 1050 Walnut St., Boulder, CO 80302, USA}

\author{Sean N. Raymond}
\affil{Laboratoire d'Astrophysique de Bordeaux, Univ. Bordeaux, CNRS, B18N, all{\'e}e Geoffroy Saint-Hilaire, 33615 Pessac, France }



\begin{abstract}

The main asteroid belt (MB) is low in mass but dynamically excited. Here we propose a new mechanism to excite the MB during the giant planet ('Nice model') instability, which is expected to have featured repeated close encounters between Jupiter and one or more ice giants ('Jumping Jupiter' -- JJ).  We show that, when Jupiter temporarily reaches a high enough level of excitation, both in eccentricity and inclination it induces strong forced vectors of eccentricity and inclination across the MB region. Because during the JJ instability Jupiter's orbit `jumps' around, the forced vectors keep changing both in magnitude and phase throughout the whole MB region. The entire cold primordial MB is thus excited as a natural outcome of the JJ instability.  The level of such an excitation, however, is typically larger than the current orbital excitation observed in the MB. We show that the subsequent evolution of the Solar System is capable of reshaping the resultant  over-excited MB to its present day orbital state, and that a strong mass depletion ($\sim$90$\%$) is associated to the JJ instability phase and its subsequent evolution throughout the age of the Solar System.

\end{abstract}



\section{Introduction}

The present-day asteroid main belt (MB) presents a challenge for theories of planet formation.  Orbits within the MB have eccentricities ranging from 0 to $\sim$0.4 and orbital inclinations from 0 to more than 20 degrees.  This high level of excitation is hard to reconcile with the presumably cold initial orbits of all planetesimals in the proto-planetary disk (including the MB asteroids and primordial trans-Neptunian objects) -- with eccentricities and inclinations near zero --  as well as the cold orbits of the terrestrial and giant planets.  The MB is low in mass, containing a total of just $\sim 5 \times 10^{-4} \mearth$~\citep{demeo2013}.  This is far less (100--1000 times less) than the few Earth-masses expected if the MB region were part of a disk with a smooth radial surface density gradient \citep[e.g.,][]{hayashi1981,bitsch15}.  Finally, the belt shows broad compositional diversity but is dominated by two prominent classes: the S-types in the inner MB and C-types in the outer MB, albeit with significant overlap~\citep{gradie82,demeo2013,demeo2014}.  Addressing these constraints within a self-consistent framework of terrestrial- and giant-planet formation is an imposing theoretical challenge \citep[e.g.,][]{morbidelli2007,hansen2009,walsh2011,izidoro2015,levison2015a,levison2015b,walsh2016,raymond2017a,raymond2017b}.

Some theories succeeded in matching the MB's level of excitation but presented other major problems.  The `classical model' of terrestrial planet formation includes a distribution of planetary embryos extending out into the asteroid belt~\citep{chambers98,chambers01,raymond06,obrien06}, which naturally excited the surviving asteroids~\citep{petit01,chambers01b,obrien2007}.  However, the classical model has a well-known Achilles heel: it systematically results in the formation of Mars analogs almost as massive as Earth~\citep{wetherill91,raymond2009,morishima10}, and very often leaves planetary embryos surviving in the belt, which is not consistent with the current observations of the MB \citep{raymond2009}. 

Sweeping secular resonances during planetesimal-driven migration of Jupiter and Saturn was also proposed to excite the MB \citep{minton2011,lykawka2013}.  However, this model requires a fast migration for Saturn ($\dot{a}\sim$ 4 au/Myr in the case of an initially cold MB or $\dot{a}\sim$ 0.8 au/Myr for an initially hot MB) with Jupiter fixed at $\sim$5.2 au \citep{minton2011}. According to \citet{morbidelli2010}, a more realistic time scale of migration for Jupiter and Saturn embedded in a planetesimal disk should be $\tau \sim$ 5 My. Planet migration on this timescale would result in a MB incompatible with that currently observed, where one would obtain an inner belt with a larger fractional number of asteroids in high inclined orbits than is observed today \citep{morbidelli2010,walshmorby2011,toliou2016}.

Rather than a smooth giant planet migration, an early 'Nice model' planetary instability has the potential to explain the asteroid belt's orbital structure. In the Nice model, the giant planets formed in a more compact and more circular/coplanar configuration than their current one, and achieved their current configuration after a phase of dynamical instability after gas dispersal. \citet{clement2018} showed that the instability can produce sufficient excitation and mass depletion exterior to $\sim$1.5 au to explain the small mass of Mars. While promising in terms of solving the small Mars problem, the simulations of \citet{clement2018} did not have high enough resolution to fully populate the MB. Rather, while they did provide a decent match to the MB they were forced to co-add many simulations to produce a model belt \citep[Figure 6 in][]{clement2018}.

The Grand Tack model \citep[GT;][]{walsh2011} was the first model to match the inner Solar System in a single evolutionary scenario. In the GT, Jupiter is assumed to have formed beyond the snow line and migrated inward via planet-gas disk interactions~\citep[e.g.][]{kley12,baruteau14}.  Meanwhile, Saturn grew and migrated inward towards Jupiter~\citep{masset03}.  When Saturn caught up with Jupiter the planets became locked in either mutual 3:2 or 2:1 mean motion resonance~\citep[MMR;][]{masset01,morby07b,pierens08,pierens11,pierens14}.  At this point the planets' direction of migration was reversed \citep[this happened when Jupiter was at around 1.5--2 au,][]{brasser2016} and both planets migrated outwards until the gas in the disk dissipated, reaching their pre-instability location \citep{nesvorny2012,deienno2017}. 

Within the framework of the GT model, Jupiter and Saturn's excursion into the terrestrial and MB region has several implications. It confines the distribution of most solid material within $\sim$1 au from the Sun, explaining why Mars accretion stopped and the planet remained small \citep{wetherill1978,hansen2009}. The excursion of Jupiter through the asteroid belt can also explain how the S-type and C-type asteroids were implanted into the MB region, with S-types originating interior to Jupiter's original orbit and the C-types farther out~\citep{walsh12}. The resulting MB population is dynamically excited, the two asteroid types are partially mixed and the total mass, is just a few times the current one.  In fact, the MB after the GT is over-excited compared with the present-day belt, but \citet{deienno2016} showed that the subsequent $\sim$4.5 Gy evolution of the Solar System naturally erodes the over-excited component so that the final distribution matches the present-day MB. A caveat, however is that the inclination distribution out of the GT simulations should be confined within $\sim$20$\degr$ in order to reproduce the current ratio in the number of asteroids above and below the $\nu_6$ secular resonance.

Despite the GT's success the scenario remains controversial \citep[see][for a critical review]{raymond2014}.  The key uncertainty is related to the outward migration mechanism of the planets, which has not been validated when gas accretion onto the giant planets is taken into account in a self-consistent way \citep{dangelo2012}.  Indeed, in an isothermal disk with Jupiter and Saturn in 3:2 MMR the planets only migrate outward for Jupiter-to-Saturn mass ratios between roughly 2 and 4~\citep{morby07b}. Of course, this ratio was evolving while the planets were accreting gas, with a direct feedback between the giant planets' growth and migration.

Another class of models invokes that the asteroid belt had a low mass from the very beginning \citep{izidoro2015,levison2015b,ogihara2015,moriaty2015,drazkowska16}. 
These models argue that the drift of small particles due to aerodynamics drag could have concentrated material near 1 au, leaving little mass in the asteroid belt. Starting from a steep enough radial surface density distribution of solid material,   \citet{izidoro2015} was indeed able to build terrestrail planets similar to the real one, with a large Earth/Mars mass-ratio, but, they could not explain the orbital excitation of the MB (which remained too dynamically cold) nor its taxonomical mixture.  However, additional mechanisms have been proposed that could reconcile the low-mass asteroid belt model with the present-day MB and provide a viable alternative to the GT model.
\citet{raymond2017a,raymond2017b} showed that even if the MB was originally empty or almost empty (compatible with the Low Mass Asteroid Belt scenario), the growth of Jupiter and Saturn during the gas disk phase naturally implants scattered primordial planetesimals into the MB region.  Planetesimals from the Jupiter-Saturn region and beyond are scattered inward during the giant planets' growth and implanted into the belt under the action of aerodynamic gas-drag \citep{raymond2017a}. They could correspond to the C-type asteroids that we observe today. In addition, planetesimals scattered outward from the terrestrial planet-forming region, due to their interaction with rogue planetary embryos, can be implanted onto main belt orbits by resonant interactions with Jupiter \citep{bottke06,raymond2017b}. They could correspond to the S-type asteroids.

While this may solve the problem of the MB's taxonomical mixture, a problem persists. The dynamical state of the MB is still cold, mainly because the implantation of C-type asteroid occur via gas-drag damping, such that all but the largest ones (D $= 1000 ~\rm km$) end up with orbital eccentricity and inclinations near zero \citep[see Fig. 3 in][]{raymond2017a}. One proposed solution is the chaotic excitation model described by \citet{izidoro2016}. In this model, Jupiter and Saturn are initially in mean motion resonance (as predicted by migration models) but are not very close to the resonance center, so that they have some chaotic motion on secular timescales. It remains to be demonstrated whether such a specific configuration is consistent with migration models of resonant capture. Indeed, the chaotic excitation has only been demonstrated when Jupiter and Saturn are initially locked in their mutual 2:1 MMR and for specific configurations inside this resonance.
However, according to \cite{nesvorny2012} and \citet{deienno2017}, it is more likely to reconstruct the orbital architecture of the outer Solar System with  a giant planet instability if Jupiter and Saturn  were initialy in the 3:2 MMR. 

The goal of this paper is to better understand the evolution and dynamical excitation of a cold primordial MB during the giant planet dynamical instability.  Our study starts from a best guess for the initial configuration of the giant planets proposed by \citet{deienno2017}\footnote{Similar results could be expected from the evolution proposed by \citet{gomes2018}, due to the fact that what matters is the evolution of Jupiter durring the JJ-instability phase and not the one of Neptune as in \citet{nesvorny2015}.}, with Jupiter and Saturn initially locked in their 3:2 MMR. We show how the evolution of Jupiter is the key for the understanding of both the excitation of the MB and the chaotic evolution by \citet{izidoro2016}, when starting from a resonant configuration of the giant planets. 
Our approach is similar to that of \citet{clement2018} but with two main differences. First, we start with a low-mass asteroid belt with the goal of demonstrating dynamical excitation with little focus on mass depletion. Second, we consider enough particles in order to assess the final orbital distribution in the asteroid belt with good statistics. 

This study is not meant to disprove neither the GT nor the chaotic excitation model. Rather, we describe a new mechanism for exciting the asteroid belt starting from a dynamically cold, low-mass setup.  Our mechanism is a by-product of the giant planet instability and, even though the belt may be temporarily over-excited, subsequent dynamical evolution brings it a state consistent with the present-day belt. It is also out of the scope of the present paper to try to provide an exact match of the present day MB, which would demand a prohibitive number of simulations and testing over a too large number of parameters.

Our paper is structured as follows.  In section \ref{sec2} we present our instability model and the effects that it has on a primordial cold MB. 
In section \ref{sec3} we discuss the direct effect that each planet has upon the MB. 
Section \ref{sec4} is devoted to understanding the mechanism of excitation working behind our results. 
In section \ref{sec5} we discuss the implications of our results for the time of the planetary instability, 
by considering the effects on the excitation of terrestrial planets. 
We compare the MB from our simulations with the present day asteroid main belt in section \ref{sec6}, where we also discuss the constraint on the initial mass of the asteroid belt.
Lastly, section \ref{sec7} concludes the paper.

\section{planetary instability and the excitation of the mb}\label{sec2}

We designed a set of simulations to test whether the MB could have been excited by the giant planet instability.  

We first performed simulations of the instability with the goal of finding cases that matched all of the constraints listed in \citep{nesvorny2012,deienno2017}. Simulations started from the five giant planet system initial configuration proposed by \citet{deienno2017}, with $a_J \sim$ 5.4 au and more distant planets locked in a resonant chain, with MMRs between adjacent planets of 3:2; 3:2; 2:1; and 3:2. Uranus, Neptune and the additional planet are assumed to have a mass of $\sim$15 Earth masses, and Jupiter and Saturn their current masses. In this configuration, all planets were initially in quasi-circular and planar orbits, as predicted by hydrodynamical simulations of planet migration, that explain the formation of the aforementioned resonant chain \citep{morbidelli2007}. We assume the existence of a planetesimal disk composed of 35 Earth masses equally divided within 1000 planetesimals with zero eccentricity and inclinations ranging from 0--1$\degr$. The planetesimal disk was designed to have a surface density of $1/r$, and the inner and outer edges were set to be 21 au and 30 au, respectively. 

We performed over one thousand simulations leading to planetary instability of the jumping-Jupiter (JJ) kind. These are evolutionary pathways in which Jupiter is involved in close encounters with another planet, typically a Neptune-mass planet that is ejected from the system. As a consequence of these encounters, the orbital distance between Jupiter and Saturn jumps abrumptly (hence the name  `jumping-Jupiter'). We extended most simulations to 10 My after the instability but could not control the exact timing of the instability.  The integrations were performed using the Mercury \citep{chambers1999} hybrid integrator with a time step of 0.5 years.  The outer disk planetesimals perturbed the planets' orbits but did not self-interact.

%
\begin{figure}
	\includegraphics[width=\columnwidth]{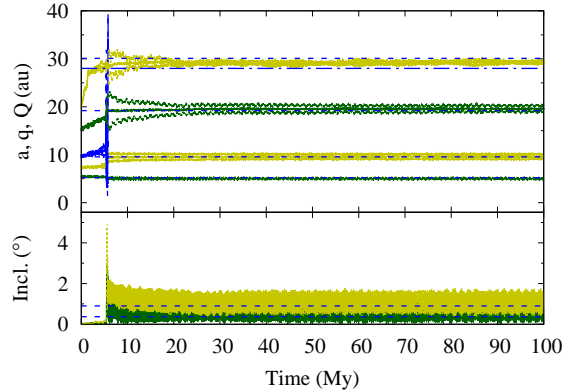}
    \caption{
Top: Evolution of the semimajor-axis, perihelion and aphelion of Jupiter, Saturn, and three ice-giant planets. The initial resonant  configuration of the planets was 3:2; 3:2; 2:1; 3:2 with Jupiter at $a_J \sim$ 5.4 au.
Bottom: Evolution of the inclination of Jupiter (green) and Saturn (yellow). 
The dashed lines in the top panel represent the current semimajor axes for the 4 Solar System's giant planets, and the dot-dashed line demarcates 28 au, were Neptune likely needs to have been at the onset of the planetary instability \citep{nesvorny2015} in order to reproduce the so-called kernel of the Kuiper belt \citep{petit2011}.
In the bottom panel the dashed lines represent the current mean inclination for Jupiter and Saturn respect to the invariable plane (the plane perpendicular with the total angular momentum of the Solar System excluding the recent proposed planet 9 \citep{batygin2016}), from \citep{nesvorny2012}, 0.37$\degr$ and 0.90$\degr$ respectively. 
}
    \label{fig1}
\end{figure}

Figure \ref{fig1} shows the most successful simulation from this batch. It matches all of the constraints presented in \citet{nesvorny2012} and \citet{deienno2017}: \\
\noindent (A) the final planetary system has four giant planets; \\
\noindent (B) the final semimajor axis of each planet is within 20$\%$ of its current value (here we got all planets within 5$\%$) and the final mean eccentricities and inclinations are no larger than 0.11 and 2$\degr$, respectively; \\
\noindent (C) the proper mode of Jupiter's eccentricity $e_{55}$ is at least half of its current value of 0.044 (see figure \ref{fig3} for C and D); \\
\noindent (D) the ratio between the orbital period of Jupiter and Saturn ($P_S/P_J$) evolves from $<$ 2.1 to $>$ 2.3 in less than 1 My; and \\
\noindent (E) Neptune migrates ahead of all other planets such as the planetary instability happens when 27 au $<~a_N~<$ 29 au, namely 28 au \citep{nesvorny2015}. 

Criterion (D) is generally invoked to guarantee the stability of the terrestrial planets in case of a late giant planet instability in the Solar System \citep{gomes2005,brasser2009,agnor2012,bottke2012}. However, it is important even in the case of an early (pre-terrestrial planet formation) instability because criterion (D) guarantees that the inner main belt is not overexcited in inclination \citep{walsh2011,toliou2016}.

Criterion (E) is imposed to make the planet evolution consistent with the current structure of the Kuiper belt \citep{nesvorny2015}.

The simulation from Fig \ref{fig1} represents a self-consistent evolution of the Solar System and we refer to it as the nominal simulation$\footnote{We did not restrict ourselves to only this simulation. Several other cases confirm the findings we will report with the nominal simulation. However, most of these additional cases do not satisfy all constraints A--E. 
In fact, finding a case as the one of figure \ref{fig1}, which satisfies all constraints, and still account for the evolution that we will discuss throughout the paper, in a small number of attempts is really difficult, given the large variety of evolutions resultant from an instability simulation. Because of that, to make short the discussion, we decided to focus in only this case. Therefore, although the results we will present may not be achieved for every instability, they certainly will every time that the instability happens as reported here.}$.

%
\begin{figure*}
	\includegraphics[width=9.cm]{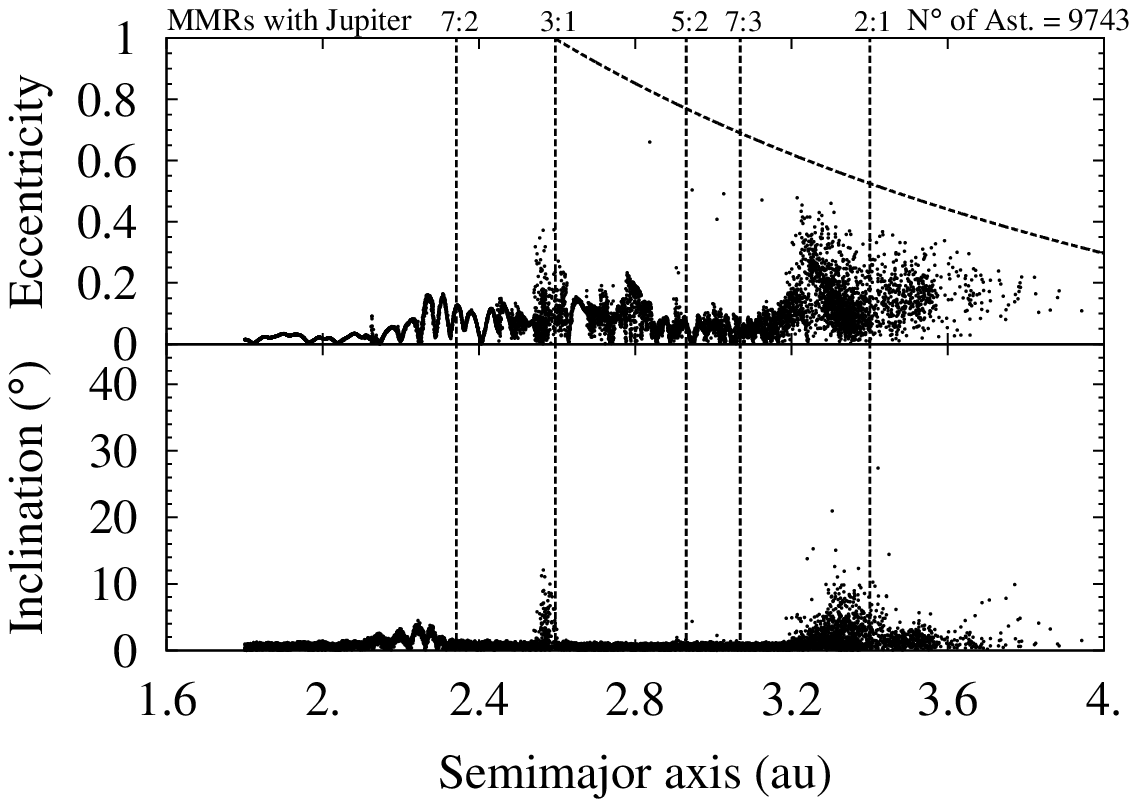}
	\includegraphics[width=9.cm]{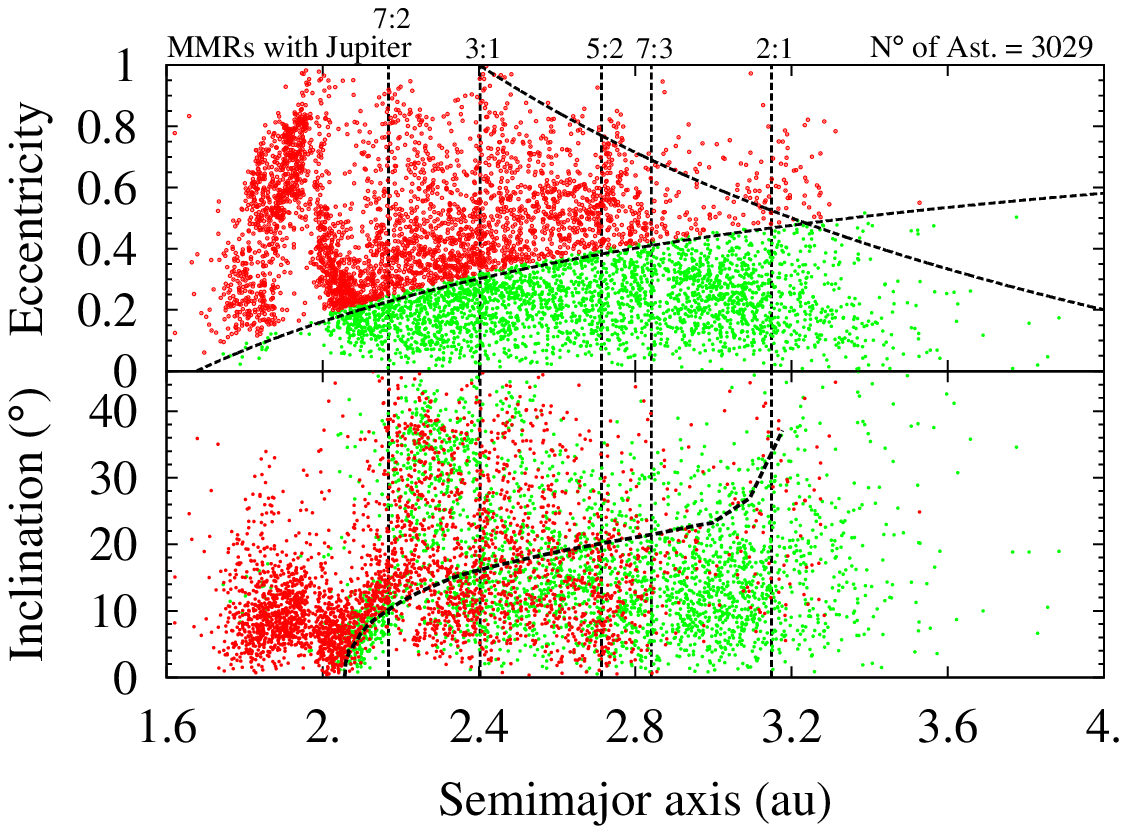}
    \caption{
Left: Eccentricity (top) and inclination (bottom) of the MB just before the planetary instability. Right: Eccentricity (top) and inclination (bottom) of the MB excited by the JJ planetary instability marked in green for eccentricities below the Mars- crossing orbit line, and in red when above it, for both top and bottom panels.
The vertical dashed lines show the MMRs between asteroids and Jupiter.
The curved dashed lines in the top panels represent the boundary of the Jupiter-crossing orbit (left and right panels), and Mars-crossing orbit (top right panel).
In the bottom right panel, the dashed curved line represents the current position of the $\nu_6$ secular resonance.
}
    \label{fig2}
\end{figure*}

Next, we ``embedded" a primordial cold asteroid belt into this nominal simulation.  To save CPU-time, we re-simulated our nominal case and recorded the orbits of Jupiter, Saturn, and the ejected extra ice-giant (hereafter $pl5$) at every 1 year output, starting from $\sim$2.5 My before the planetary instability for a total integration time of 5 My. From these outputs, we interpolated the orbits of the planets (using linear interpolation of their orbital elements) from $\sim$2.5 My before to $\sim$2.5 My after the planetary instability.  The orbits of Uranus, Neptune, and of the trans-neptunian planetesimals were not considered (but their presence is indirectly recorded in the orbital evolution of the considered planets). We included a cold MB extending from 1.8 au to 3.6 au consisting of 10,000 massless particles with eccentricities equal zero and inclinations ranging from 0-1$\degr$. The other angular elements were drawn randomly within the interval 0--360$\degr$.  We modified the Mercury integrator to include the planet's interpolated orbits in order to compute the external force that perturb the test particles, and reduced the integration time step to 10 days. The terrestrial planets were also ignored in this phase of the simulation. Although this may prevent most of the high eccentricity asteroids in the inner MB from been removed, at this point we are only interested in evaluating the level of excitation provided by the instability of the giant planets. Having the terrestrial planets included in the simulation would probably cause some depletion of the asteroids colored in red in figure \ref{fig2} (right panels). However, the terrestrial planets would not significantly modify the precession rates of the main belt asteroids, which are dominated by Jupiter \citep[see eq. 7.55 of][]{murray1999} and, as we show below, is the main driver of the excitation process. Therefore, including the terrestrial planets is not important during the instability. Not knowing whether the terrestrial planets already existed or not at that time, we prefer not to include them. Of course, they will be included in the subsequent simulation on the long term evolution of the MB in the post-excitation phase.

The MB's final level of excitation from this nominal simulation is shown in figure \ref{fig2} right panel, and the state of the MB just before the JJ instability is shown in figure \ref{fig2} left panel. The excitation is similar to those presented in \citet{deienno2016} for the GT and in \citet{izidoro2016} for the chaotic excitation and greatly exceeds the current dynamical excitation of the MB. This suggests that, whatever the excitation mechanism, an over-excitation phase for the MB seems unavoidable. However, a close look in the evolution of the MB$\footnote{An animation of the entire evolution can be found electronically at http://staff.on.br/rodney/rogerio/mb-excitation.mp4 with Jupiter (gray), Saturn (orange), and $pl5$ (green).}$ showed no path to orbital chaos prior the instability when the initial configuration of the planets were 3:2; 3:2; 2:1; 3:2. In other words, as there was no chaotic excitation before the planetary instability, unlike in \citet{izidoro2016}, a different mechanism is at play. Still, a very strong perturbation occurred during a very short window of time ($<$ 200 ky)  within the JJ instability period ($\sim$800 ky or less, figure \ref{fig3}).

As a last note, it is important to say that, although Jupiter and Saturn go beyond the 5:2 MMR (i.e. their final separation is larger than their current one), the crossing of the 5:2 MMR seems to play no important role in the entire excitation process. Thus, one results should not be an artifact of this feature in the planets' evolution. 

%
\begin{figure}
	\includegraphics[width=\columnwidth]{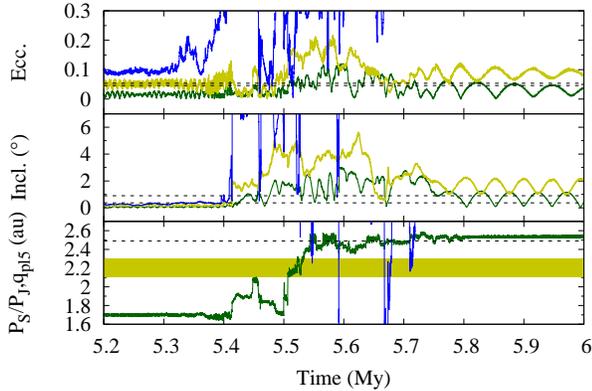}
    \caption{
Evolution of the eccentricity (top) and inclination (middle) for Jupiter (green), Saturn (yellow), and $pl5$ (blue), as well as in the bottom panel, the evolution of $P_S/P_J$ (green) and of perihelion of $pl5$ ($q_{pl5}$ -- blue).
The horizontal dashed lines in the top and middle panels represent the present day mean values of Jupiter and Saturn eccentricities and inclinations, respectively.
The yellow shaded area in the bottom panel represent 2.1 $<~P_S/P_J~<$ 2.3 where the period ratio has to spend the shortest possible time in order to avoid a strong depletion of the inner asteroid belt \citep{morbidelli2010} and the horizontal dashed line in the same panel the current value of $P_S/P_J$.
}
    \label{fig3}
\end{figure}

~

%
 \begin{figure*}
 	\includegraphics[width=6.2cm]{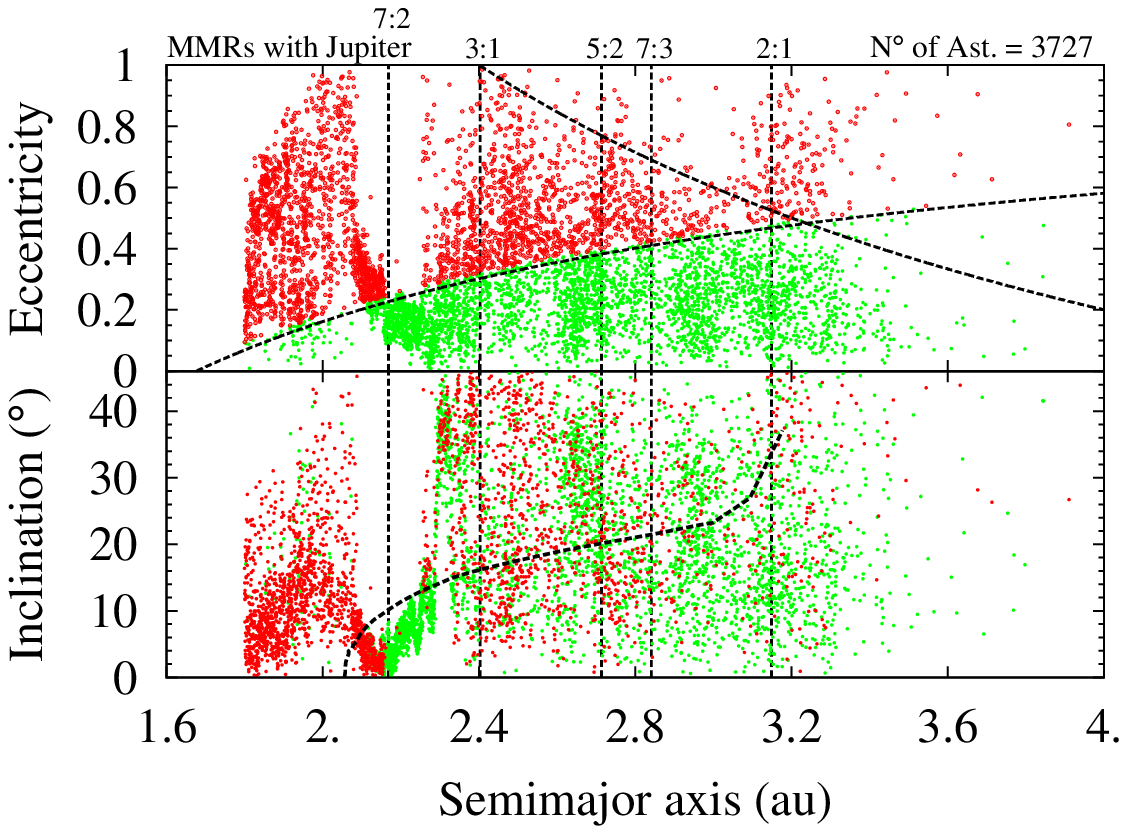}
 \hskip -.5cm
 	\includegraphics[width=6.2cm]{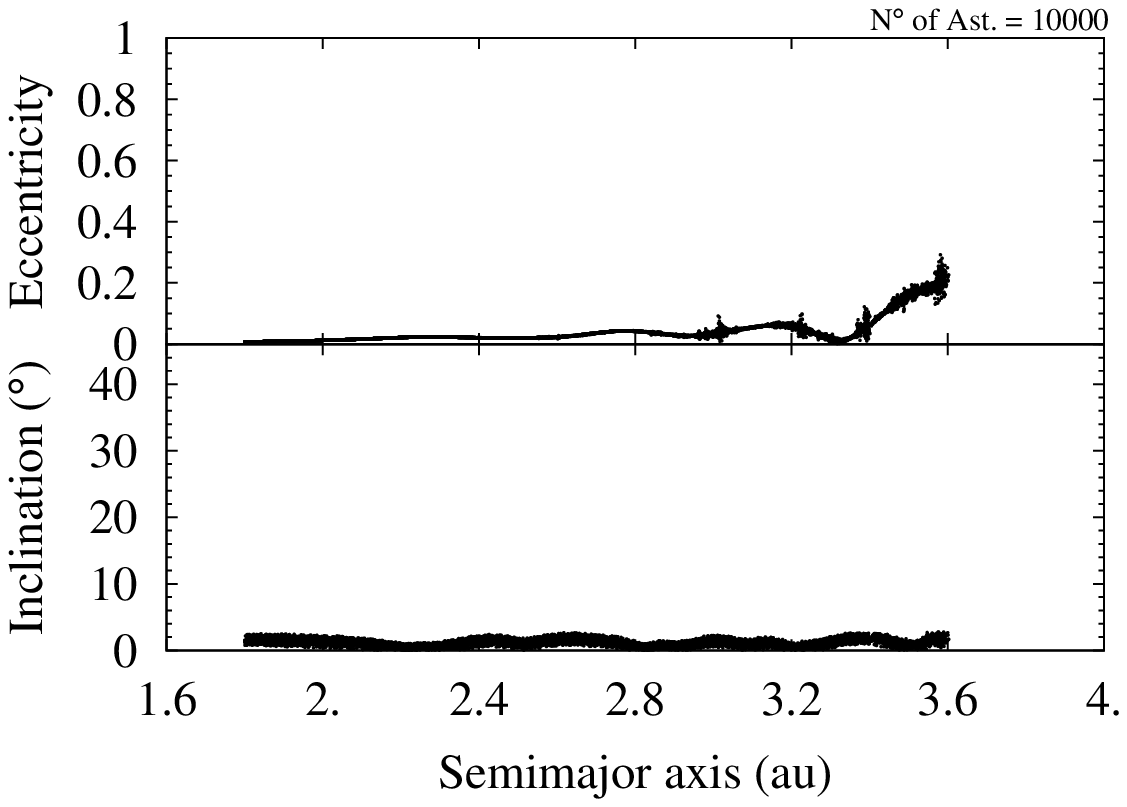}
 \hskip -.5cm
 	\includegraphics[width=6.2cm]{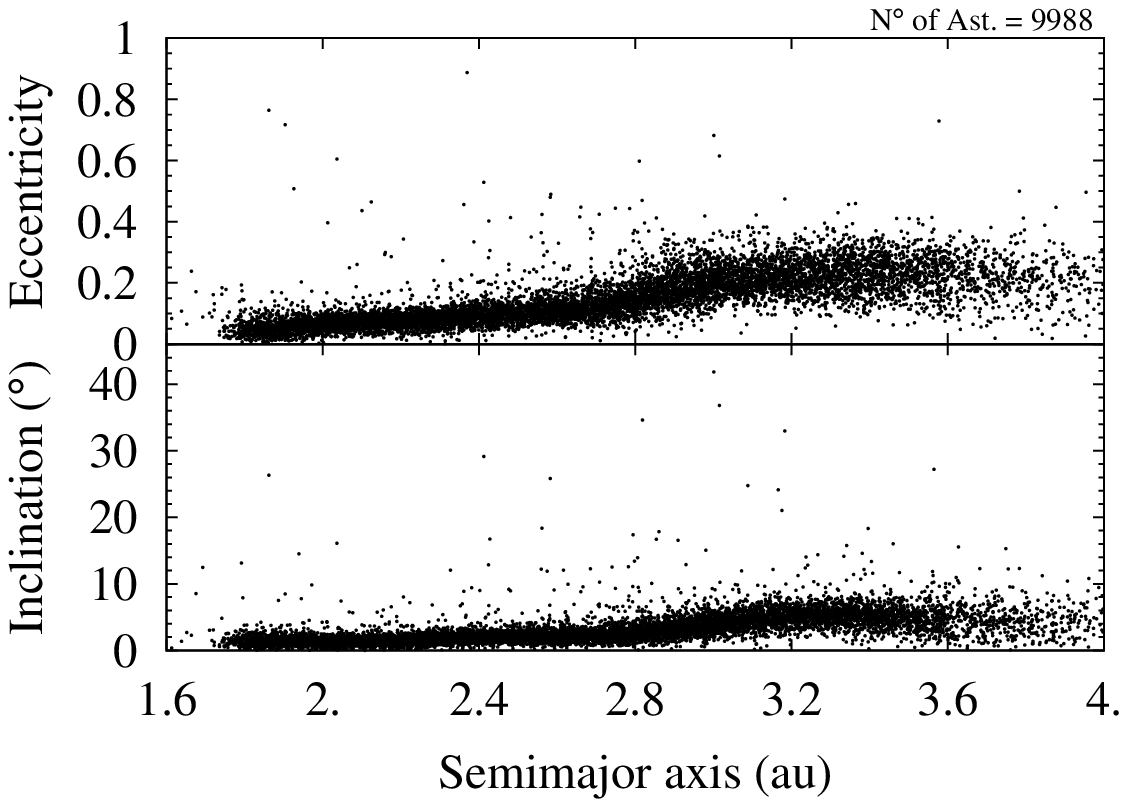}
%
%
    \caption{
Eccentricity and inclination as a function of the semimajor axis for the main belt asteroids at the end of the 800 ky simulation passing through the JJ planetary instability when considering only Jupiter (left), only Saturn (center), and only $pl5$ (right).
The vertical dashed lines in the left panels show the MMRs between asteroids and Jupiter.
The curved dashed lines in the left panels represent the boundary of the Mars- and Jupiter-crossing orbit (top) and the current position of the $\nu_6$ secular resonance (bottom).
Also in the left panels, asteroids above the Mars-crossing orbit are marked in red and those below this line in green, for both top and bottom panels.
}
    \label{fig4}
\end{figure*}

\section{The role of each planet}\label{sec3}

To better understand the MB excitation during the planetary instability, we now focus on the short window of time (800 ky) when the instability happens. 
Figure \ref{fig3} shows the evolution of the eccentricity and inclination of Jupiter (green), Saturn (yellow), and $pl5$ (blue) during that window of time, as well as the evolution of $P_S/P_J$ (green) and of the perihelion of $pl5$ ($q_{pl5}$ -- blue).
Figure \ref{fig3} shows the interval before and after the instability shown in figure \ref{fig1} in much more detail. 

Our next step was to test the individual effect of Jupiter, Saturn and $pl5$ on MB excitation.  We used the orbits of Jupiter, Saturn, and $pl5$ interpolated during the 800 ky of figure \ref{fig3} to simulate their individual effects on an initially cold MB.  The initial belt was the same as previously described, with 10,000 massless particles with $a=$ [1.8--3.6] au, $e =$ 0, incl. = [0--1]$\degr$, and $\omega, \Omega, l$ = [0--360]$\degr$.
We effectively re-ran the same MB three times but considering only one of the three planets in each simulation.
As the orbits of the planets were interpolated, we did not lose any detail in the path followed by each of them.
Of course the frequencies that they induce in the MB change by changing the number of disturbing planets, but this experiment was simply to evaluate only the direct effect caused by each planet.

Figure \ref{fig4} shows the eccentricity and inclination distributions for main belt asteroids at the end of the 800 ky simulation passing through the JJ planetary instability when considering only Jupiter (left), only Saturn (center), and only $pl5$ (right).

It is clear from figure \ref{fig4} that Jupiter is the planet responsible for exciting the MB during the planetary instability. The non-randomized eccentricity and inclination excitation between 2 au and 2.4 au in the left panel of figure \ref{fig4} is most probably related to the incorrect secular frequencies of asteroids due to the absence of the direct effect of Saturn, which displaces the location of secular resonances. Even so, the result itself is very similar to that presented in figure \ref{fig2} right panel where all three planets (Jupiter, Saturn, and $pl5$) were considered together, specially when considering the green colored asteroids below the Mars-crossing line.
Saturn alone, middle panels of figure \ref{fig4} seems to have an almost negligible direct effect upon a primordially cold MB. Surprisingly, even $pl5$, which has a high eccentricity and a perihelion distance well inside the MB for some periods during the considered time-interval ($q_{min} \sim$ 0.57 au at $t \sim$ 5.59 My, when $a \sim$ 3.15 au, $e \sim$ 0.82, and $i \sim$ 9.37$^o$ or $q_{min} \sim$ 1.26 au at $t \sim$ 5.67 My, when $a \sim$ 5.34 au, $e \sim$ 0.76, and $i \sim$ 11.66$^o$ -- figure \ref{fig3} bottom panel) does not excite the initially cold MB to the level in figure \ref{fig2} right panel. Most probably, the weakness of $pl5$ in exciting the MB is related to its large eccentricity and inclination when attaining a low pericenter distance.
Only the regions of the MB with $a~>$ 2.8 au faced some non negligible disturbance, but the eccentricities reached no more than 0.4 and inclinations barely reached 10$\degr$ (figure \ref{fig4} right panel). 
The question that remains is: why is Jupiter alone causing so much excitation, mainly in inclination?  The answer can likely offer a complementary explanation of the chaotic evolution seen in \citet{izidoro2016} and can also be related to the dispersal of asteroidal families during the planetary instability proposed by \citet{brasil2016}, although with somewhat smaller dispersion in the inclinations in their work.

In fact, the excitation of the inclination is difficult to understand. The Jupiter-Saturn system has only one secular frequency, called $s_6$, associated to the motion of their longitudes of the node. So, there is only one nodal secular resonance possible between the asteroid and the planets: $s=s_6$, aka $\nu_{16}$. In the jumping-Jupiter (JJ) the $\nu_{16}$ resonance jumps across most of the asteroid belt, instead of sweeping through it. So one should expect that the middle and outer parts of the asteroid belt remain unexcited in inclination.

\section{Mechanism of excitation}\label{sec4}

Having in mind that the inclination excitation of the MB is the key to constrain the evolution of Jupiter during the JJ phase, we closely studied the inclination evolution of selected MB test particles.
We randomly choose a few MB test particles across the MB range. 
What we found is that the response in the inclination of all selected MB test particles was related to changes in the inclination of Jupiter. 
Every time the inclination of Jupiter went up or down, the inclination of the MB test particles responded by going up or down, and also acquiring different amplitudes of oscillations. 
Such a response might be an effect of the forced inclination of Jupiter acting upon all MB test particles.

\begin{figure}
\gridline{\fig{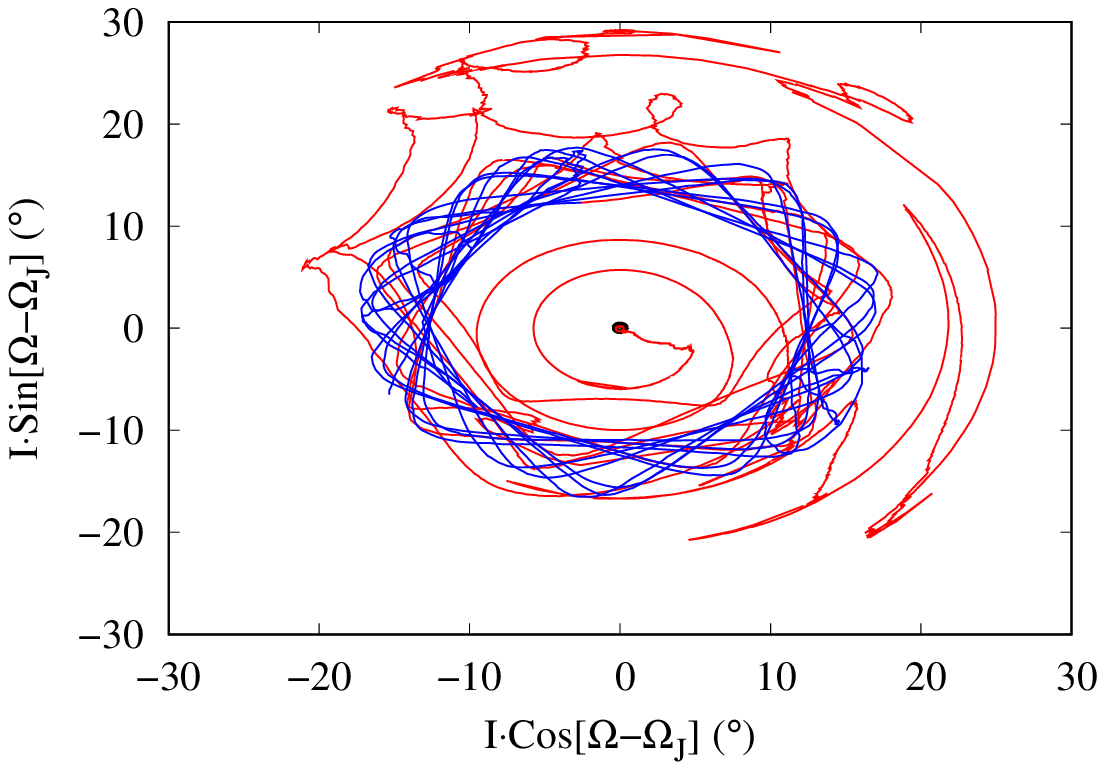}{.25\textwidth}{}
\hskip -.5cm
          \fig{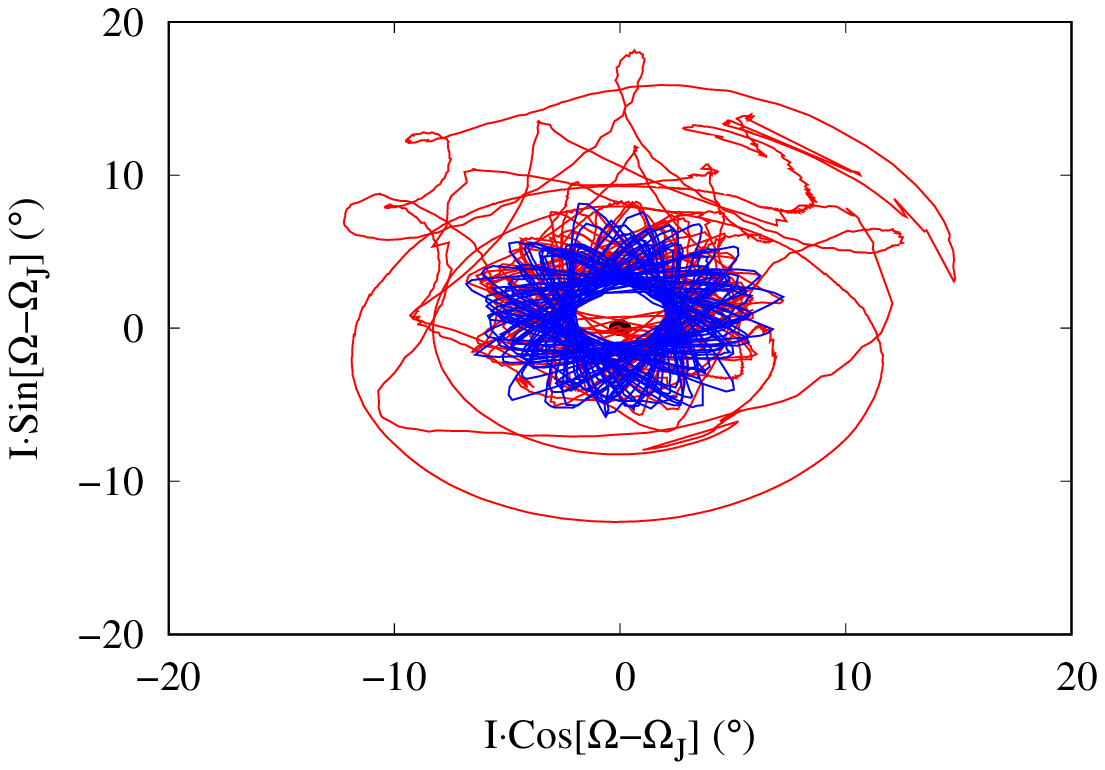}{.25\textwidth}{}}
\vskip -1.25cm
\gridline{\fig{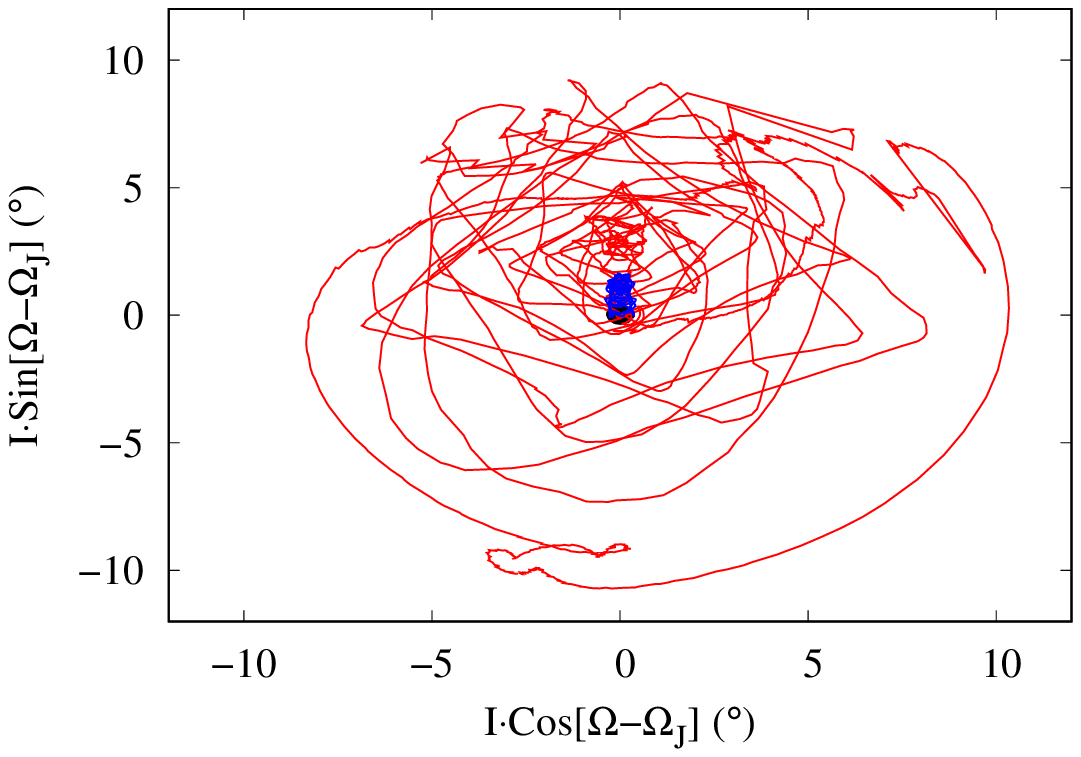}{.25\textwidth}{}
\hskip -.5cm
          \fig{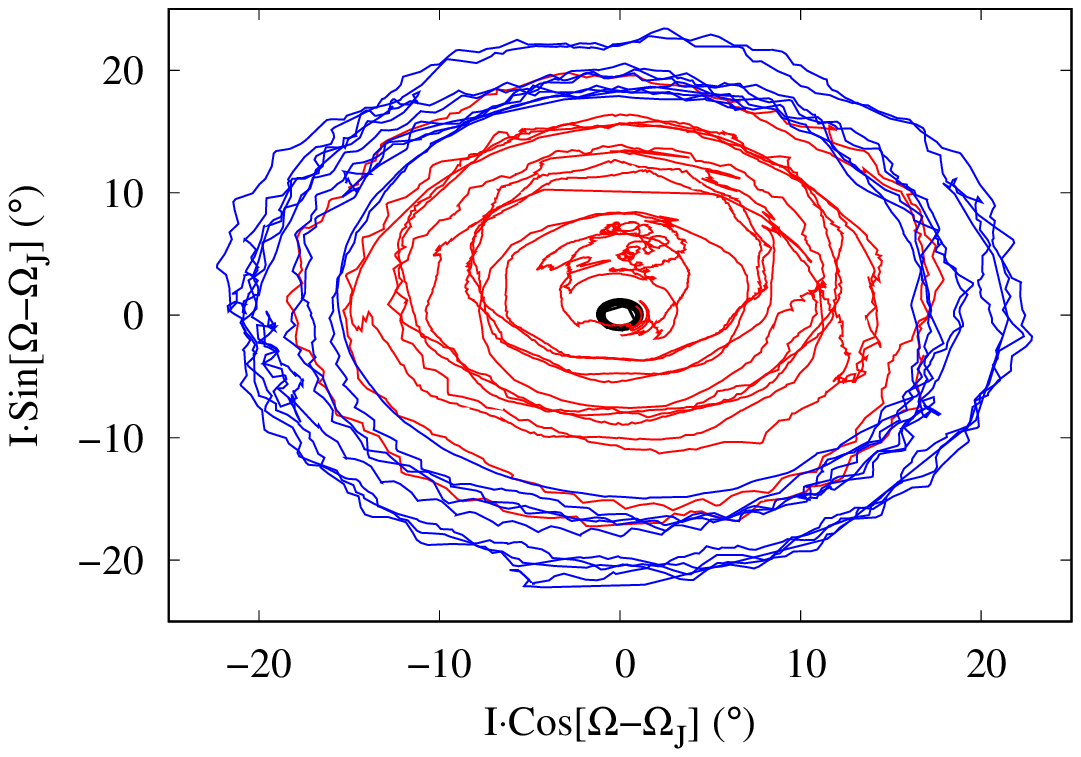}{.25\textwidth}{}}
\vskip -.75cm
    \caption{
$I\sin(\Omega-\Omega_J)~vs~I\cos(\Omega-\Omega_J)$ for four randomly selected MB test particles, for three different epochs, before (small central black circle), during (red), and after (blue) the planetary instability.
}
    \label{fig5}
\end{figure}

Figure \ref{fig5} illustrates this phenomenon. It plots $I\sin(\Omega-\Omega_J)~vs~I\cos(\Omega-\Omega_J)$ for four MB test particles during three different epochs: before, during, and after the planetary instability.

%

The different behaviors shown in these plots indicate that: 
1) in the beginning, when Jupiter has a well behaved quasi-circular and planar orbit, the forced inclination vector ($I_{forced}$) felt by the MB test particles is small (small central black circle in the panels of figure \ref{fig5});  
2) at the onset of the instability, when Jupiter gets some inclination, it induces an instantaneous large forced inclination vector on MB test particles. 
Then, as Jupiter jumps around, the magnitude and phase of the $I_{forced}$ vector changes. 
Thus, at different locations within the MB, different MB test particles, with different longitude of ascending node, can undergo different amplitudes of oscillations. 
For this reason, the final inclination excitation is not uniform (i.e. with all particles within a given semimajor axis reaching the same inclination). Instead, particles inclinations are randomized, filling all the inclination range (red curve in the panels of figure \ref{fig5});
3) once the planetary instability phase ends and Jupiter returns to a more well behaved, quasi-circular and low inclined orbit\footnote{Although in the end of the 800 ky simulation, figure \ref{fig3} middle panel, Jupiter's inclination is higher than its current value and with large oscillations, figure \ref{fig1} bottom panel shows that such inclination will continue damping over the next 100 My, matching perfectly the constrains imposed and well representing Jupiter's current inclination.}, the MB test particles return to face a constant magnitude of the inclination forced vector, and because of that, their proper inclinations are frozen at the values acquired during the JJ instability phase (blue curves in the panels of figure \ref{fig5}).

This mechanism is consistent with the chaotic excitation of \citet{izidoro2016}, although in that model this happened during the phase when Jupiter and Saturn were in their mutual 2:1 MMR, whereas here it occurs when the planets are evolving in semimajor axis under the effects of mutual close encounters. Our results therefore suggest a self-consistent evolution that leads to a pathway for a possibly chaotic excitation of the MB induced by the JJ planetary instability, which works also with Jupiter and Saturn initially in their 3:2 MMR.

Summarizing, to obtain a well excited MB from a primordially cold one all orbital elements of Jupiter have to change several times during the JJ instability phase. Jupiter must also acquire a non negligible inclination during the instability phase to excite the MB inclinations. Although in our runs that led to high enough level of inclination excitation of the MB the orbital inclination of Jupiter was $\sim$1.5$\degr$ to $\sim$3$\degr$ during the JJ instability (figure \ref{fig3}), we cannot at this point constrain the minimum inclination that Jupiter had to achieve.
This is because the forced inclination vector faced by the MB test particles does not depend directly on Jupiter's inclination, but rather on the amplitude of the $s_6$ frequency in its secular motion, the value of the latter, as well as the frequencies and associated amplitudes induced by the presence of the ice giant planets.  This becomes evident when developing an analytic theory for secular motion, along the lines of chapter 7 in \citet{murray1999}. Because in the JJ evolution the orbits of the planets keep changing, the frequencies, the secular phases and the amplitudes keep changing as well.

Finally, it is also important to discuss three other aspects: \\
($i$) as the instability depicted in this work requires a relatively long sequence of Jupiter's jumps, it is important to check that a long phase of close encounters between Jupiter and $pl5$ does not destabilize the regular satellite system of Jupiter \citep{deienno2014}. \citet{deienno2014} concluded that what is important for the stability of the satellites is not the total number of encounters, but the distance of the closet one. Following \citet{deienno2014}, the orbits of the Galilean satellites can be profoundly affected if the encounter distance (d$_{enc}$) between Jupiter and $pl5$ is d$_{enc}$ $<$ 0.03 au. They found that, 0.03 au $<$ d$_{enc}$ $<$ 0.05 au causes only small variations in the orbital elements of the Galilean satellites, and that d$_{enc}$ $>$ 0.05 au leave the satellite system essentially undisturbed. In our nominal simulation we never get d$_{enc}$ $<$ 0.05 au. To be more precise, the three closest encounters we measured were d$_{enc} \sim$ 0.05 au at t $\sim$ 5.46 My, d$_{enc} \sim$ 0.06 au at t $\sim$ 5.59 My, and d$_{enc} \sim$ 0.08 au at t $\sim$ 5.60 My. Therefore, our simulated evolution should not put at risk the orbital structure of the Galilean satellites.\\ 
($ii$) Although very similar, this work presents some differences from those of \citet{morbidelli2010} and \citet{brasil2016} for what concerns the excitation achieved for the orbital inclination of the MB. The main difference between these works is that we have a high time-resolved orbital evolution for Jupiter, which is also involved in a much larger series of encounters \citep[a case well prefered by][for the capture of irregular satellites]{nesvorny2014}. This leads Jupiter to change a lot in $a, e, i$, and with that the forced inclination vector. This difference alone is enough to make this case different from \citet{morbidelli2010}, which, without $pl5$ could not have this richness in stochastic events in Jupiter's evolution. In our understanding, the fact that Jupiter acquires some $\sim$1.5$\degr$ to $\sim$3$\degr$ inclination when receiving several semimajor axis kicks during the instability phase is what makes the results in this work different from those of \citet{brasil2016}. Recall that, although considering only families and not the entire MB, similar to this work, \citet{brasil2016} claim that the dispersion of the primordial families beyond recognition due to the planetary perturbations, occur particularly in inclination. However, as \citet{morbidelli2010}, they pointed out that the dispersion in inclination decreases with the increase of the semimajor axis within the MB. We argue that, with a more jumpy orbital evolution for Jupiter, acquiring a higher excitation in orbital inclination, \citet{brasil2016} would find similar results to those presented here. Therefore, \citet{brasil2016} and this work show that both a strong and a weak inclination excitation of the MB are possible, depending on the exact evolution of Jupiter's orbit. \\
($iii$) until now, we only looked at the influence of the JJ instability on the MB, from $\sim$1.8 au to $\sim$3.6 au.
However, it is possible that some large values of forced eccentricity and inclination vectors influence the terrestrial planet region as well.
The next section will address this issue.

\section{Implication for the time of the planetary instability}\label{sec5}

Here we address the dynamical excitation of terrestrial bodies during our nominal JJ planetary instability.

We once again make use of the 800 ky instability phase of our nominal simulation shown in figure \ref{fig3}.
We consider three different situations:
\begin{enumerate}
\item an early instability scenario, 10,000 massless test particles spread from 0.5 au to 4 au, with $e=$ 0, Incl.= [0--1]$\degr$, and $\omega,\Omega,l=$ [0--360]$\degr$.
\item a late instability scenario, with no MB test particles but including the fully formed Venus, Earth and Mars. In this case, we are only interested in how the terrestrial planets would respond to the JJ planetary instability and do not address the excitation of the MB because it was done in the previous sections. We consider all planets initially with their current orbits, with inclinations referred to the invariable plane \citep{souami2012}, as defined in the caption of figure \ref{fig1}.
\item a late instability scenario as in (2) but assuming all terrestrial planets from Venus to Mars to have originally circular and planar orbits.
\end{enumerate}

By looking at the evolution in the terrestrial planet region, our goal is to address whether the planetary instability is more likely to have happened early \citep{kaib2016,nesvorny2017,morbidelli2018,clement2018} or late \citep{gomes2005,bottke2012,brasser2013,roig2016} in the evolution of the Solar System.

In other to account for the perturbations of Jupiter, Saturn, and $pl5$ on Venus, Earth, and Mars, we updated our interpolation code of section \ref{sec2} so that terrestrial planets could also feel the external interpolated disturbers.
So, for (2) and (3) above, all terrestrial planets interact with each other and are perturbed by Jupiter, Saturn, and $pl5$.
As for (1), we continue using the code described in sect. \ref{sec2}, where test particles do not interact among themselves and only feel the perturbation of the interpolated planets.

\begin{figure}
  \fig{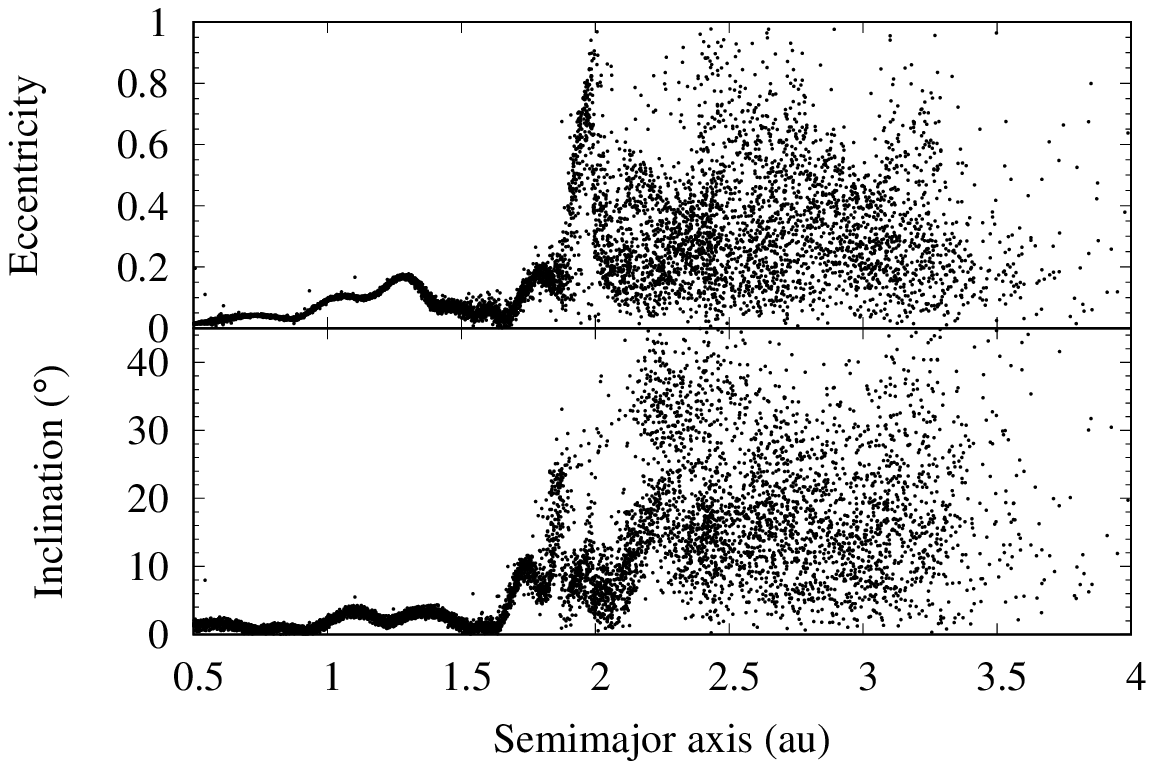}{.45\textwidth}{}
\vskip -1.25cm
   \fig{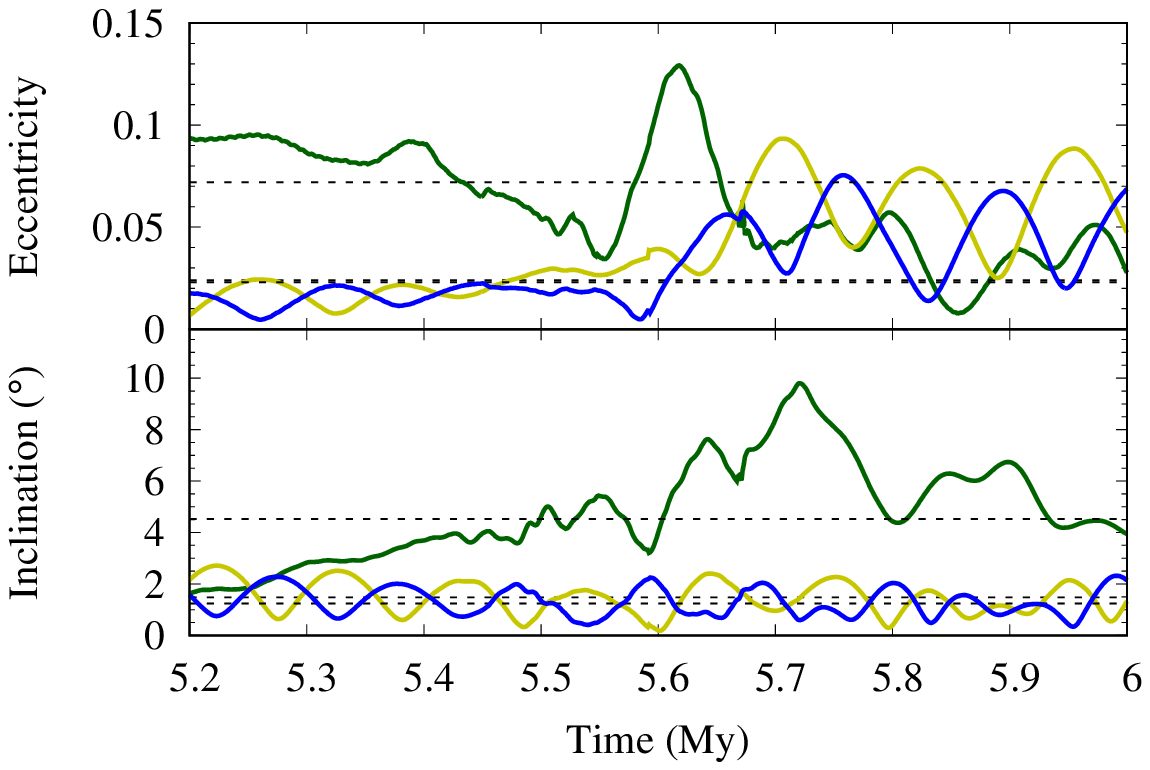}{.45\textwidth}{}
\vskip -1.25cm
  \fig{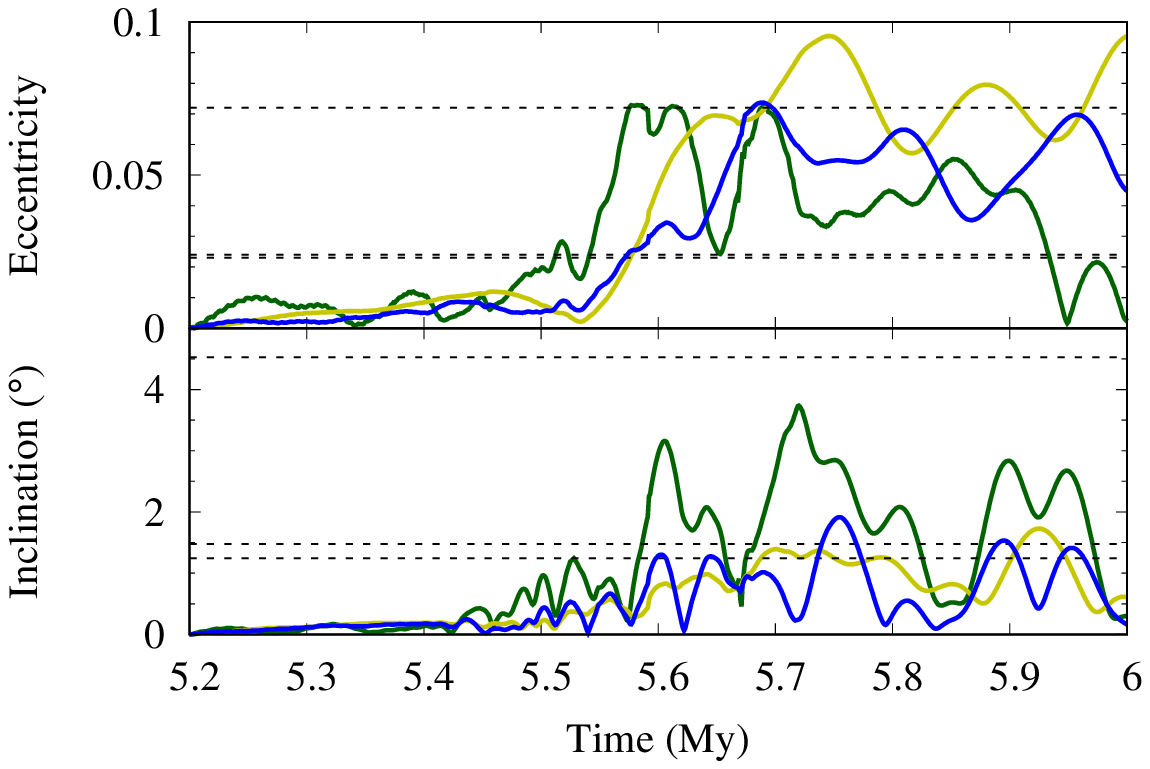}{.45\textwidth}{}
\vskip -1.cm
\caption{
Top: Distribution of eccentricity and inclination of 10,000 massless test particles initially spread from 0.5 au to 4 au, with $e=$ 0, Incl.= [0--1]
$\degr$, and $\omega,\Omega,l=$ [0--360]$\degr$ after 800ky instability simulation of figure \ref{fig3}. 
Middle and Bottom: Evolution of the eccentricity and inclination of Venus (yellow), Earth (blue), and Mars (green) under the effects of the planetary instability shown in figure \ref{fig3}.
Middle: all planets initially with their current orbits.
Bottom: all planets initially with circular and planar orbits.
Dotted lines in the middle and bottom panels represent the real mean value of eccentricity and inclination for these three planets. The real mean values were evaluated considering an evolution with all planets from Venus to Saturn interacting over a period of 10 My. The inclination values are reffered to the invariable plane \citep{souami2012}, as defined in the caption of figure \ref{fig1}.
}
    \label{fig6}
\end{figure}

Figure \ref{fig6} (top panels) show the outcome of experiment (1) after 800 kyr. 
The overall level of excitation in eccentricity and inclination as a response to the JJ instability appears to be much less effective in the terrestrial planet region. Although figure \ref{fig6} (top panels) present some non-negligible values ($\sim$0.2 of eccentricity between 1-1.5 au), the overall result is consistent with the findings described by \citet{clement2018}, where the region within $a<$ 1.8 au was much less disturbed than the region for $a>$ 1.8 au.

%

Figure \ref{fig6} middle and bottom panels, on the other hand, show a very different behavior for the already-formed terrestrial planets. 
The evolution of Venus (yellow), Earth (blue), and Mars (green), for both initial configurations, are strongly disturbed, and the terrestrial planets end up with orbits significantly more excited than their present orbits in eccentricity (the real mean values of the present orbits of these planets for both eccentricity and inclination are also shown in figure \ref{fig6} middle and bottom panels for comparison). 
Instead, the final orbital inclinations are similar to the real ones, particularly when the terrestrial planets start from their present day orbits (see figure \ref{fig6} middle and bottom panels).  
Finally, as commonly used to quantify the terrestrial planetary system orbital excitement, we considered the angular momentum deficit metrics \citep[AMD --][]{laskar1997}, in the form presented by \citet{chambers01}. The current AMD of the terrestrial planets from Mercury to Earth is $\sim$0.0014. The same value considering only Venus, Earth and Mars is AMD$_{VEM}$$\sim$0.0009. As for comparison, because we are not considering Mercury in our simulations, we should use the second value. The AMD for the simulation of the middle panel of the figure \ref{fig6} is, at the end of the simulation $\sim$0.0024 ($\sim$2.67$\times$AMD$_{VEM}$), and after taking an average on the orbital $a,e,i$ over the last 200 ky for all three planets $\sim$0.0018 ($\sim$2.00$\times$AMD$_{VEM}$). Similarly, the AMD for the simulation of the bottom panel of the figure \ref{fig6} is, at the end of the simulation $\sim$0.0023 ($\sim$2.55$\times$AMD$_{VEM}$), and after taking an average on the orbital $a,e,i$ over the last 200 ky for all three planets $\sim$0.0019 ($\sim$2.11$\times$AMD$_{VEM}$). 

Of course one could also argue that, in the early instability scenario, after the dissipation of the solar gas nebula, the system of planetary embryos precursor of the terrestrial planets could be disturbed in a similar manner as the fully formed terrestrial planets in the case of a late instability. In fact,  \citet{clement2018} found values of AMD 2--3 times the current value. 
However, dynamical friction from collisionally-generated fragments  (not considered in \citet{clement2018}) could potentially damp the final orbits of the terrestrial planets to match the present-day inner Solar System \citep{chambers2013,walsh2016}.
Instead, if the instability happened late, well after terrestrial planet formation, the presence of a remnant planetesimal disk and collisionally-generated fragments are unlikely and therefore the simulations presented here argue that the terrestrial planets would acquire orbits too excited in eccentricities. The same result was presented by \citet{kaib2016}.
On the other hand, \citet{roig2016} obtained good final orbits for terrestrial planets with late giant planet instabilities, while also  explaining the excitation of the orbit of Mercury. 
Therefore, it is important to have in mind that there are some evolutions during the giant planet instability that make this possible, even though the one presented here is not among them.

On this issue, some new constraints should be mentioned. \citet{marty2017} showed that about 20\% of the Xe in the Earth atmosphere should have a cometary origin. However, there is no trace of cometary Xe in the Earth's interior \citep{caracausi2016}. This suggests that the cometary bombardment (presumably associated with the giant planet instability) occurred after the formation of the terrestrial crust. Because Earth's formation and mantle crystallization took about 60-100My \citep{kleine2009}, this would imply that the instability did not occur before this time. Perhaps, enough planetesimals (possible collisional debris from debris of the moon forming event \citep{bottke06}) were still present in the terrestrial planet region in the early aftermath of Earth-formation (in fact the Earth accreted about 0.5\% of its mass from them, the so-called Late Veneer), to damp the over-excitation of the terrestrial planets' orbits. This remains to be shown.

\section{Subsequent evolution of the MB}\label{sec6}

We now extend the evolution of the excited MB presented in figure \ref{fig2} throughout the age of the Solar System and then compare it to the current MB. 

For simplicity we consider the asteroid distribution shown in figure \ref{fig2} right panel as our initial distribution, even though it was obtained in a simulation without terrestrial planets.
Following \citet{deienno2016}, we then plugged in the simulation all planets from Venus to Saturn in their current orbits and evolve the MB of figure \ref{fig2} right panel for additional $\sim$4 Gy \citep[with no interpolation, using once again the Mercury package,][ in the hybrid option with a time step of 10 days]{chambers1999}.
Because in the simulation performed to get figure \ref{fig2} the final semimajor axes of both Jupiter and Saturn were within 5\% of their present values, and their final eccentricities and inclinations were also close to their present values, with most of the asteroids not in mean resonances with them, we do not expect that changing instantaneously the planetary system causes a relevant perturbation in the asteroids' distribution. On the contrary, we believe this to be the best way to continue the evolution over the Solar System age, in order to be able to compare the final distribution of the asteroids with the current distribution.
 
%
\begin{figure*}
	\includegraphics[width=18.cm]{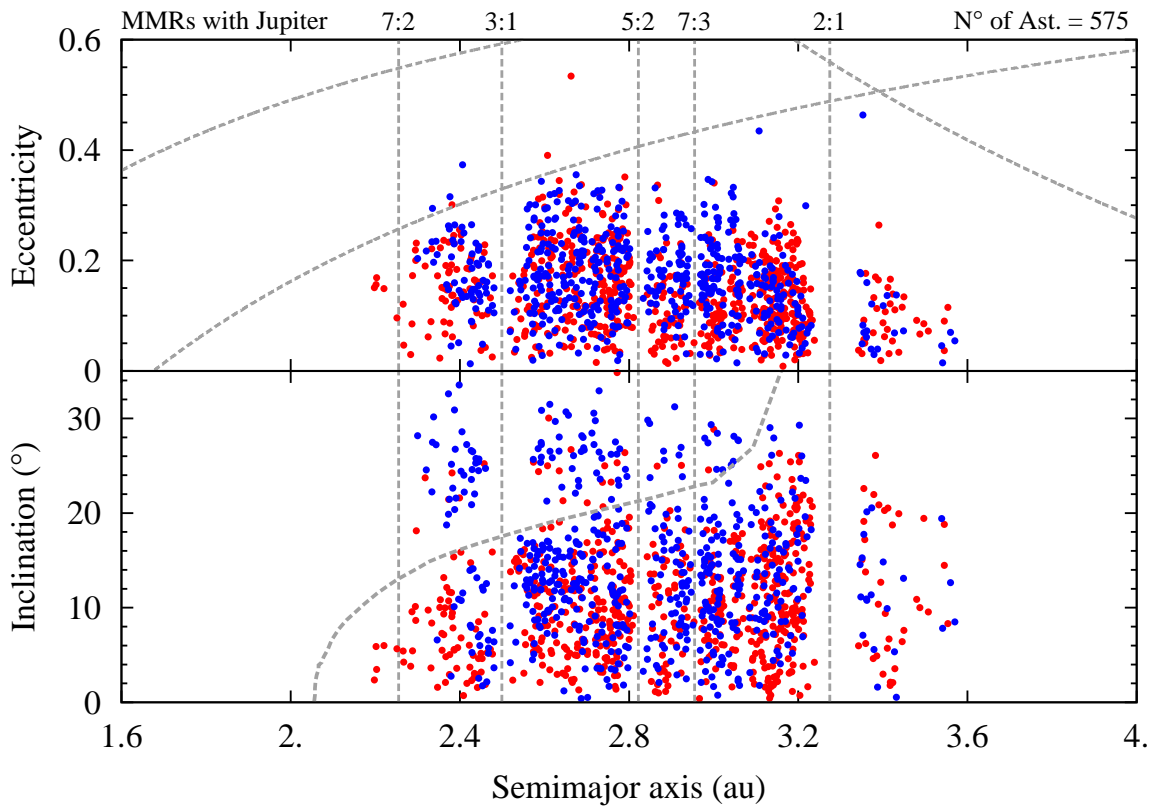}
    \caption{
Comparison between the eccentricity (top) and inclination (bottom) as a function of the semimajor axis of the MB resulting from our 4 Gy extended simulation (blue) to the present day MB (red) with all H $<$ 10.
The vertical dashed lines show the MMRs between asteroids and Jupiter.
The curved dashed lines in the top panel represents the boundary of the Earth-, Mars-, and Jupiter-crossing orbits, from left to right.
The location of the $\nu_6$ resonance is also shown in the bottom planel for reference.
}
    \label{fig7}
\end{figure*}

Figure \ref{fig7} shows a comparison between the orbital ($a,e,i$) distribution of the MB resulting from our 4 Gy extended simulation\footnote{ An animation of the entire evolution can be found electronically at http://staff.on.br/rodney/rogerio/mb-evo.mp4} (blue) to the present day MB (red) for real asteroids with\footnote{From http://www.minorplanetcenter.org/iau/MPCORB.html} H $<$ 10.

It is clear that our simulated MB provides a good match to the real one.
Of course there are a few differences. 
The major difference that should be note is in the number of objects above the $\nu_6$ resonance compared to the number of objects below the $\nu_6$ resonance for $a<$ 2.5 au, which is much larger in our simulated MB (with a ratio of objects above/below $\sim$1.3) than in the current MB ($\sim$0.09). This is similar to what \citet{deienno2016} found from their simulations of evolution and reshape of the GT's MB ($\sim$1.2). The ratio we found is nevertheless a lot better than the ones presented in \citet{walshmorby2011} ($\sim$5.2, as a result of smooth migration for Jupiter and Saturn) and \citet{clement2018} ($\sim$2.24, within JJ instability scenarios that do not take criterion D too seriously). If we exclude the asteroids with inclinations above 20$\degr$ at the end of the JJ evolution the final ratio would be $\sim$0.07.  
This suggests that the asteroid inclination excitation in the region within $a<$ 2.5 au was in reality a bit weaker than in our nominal JJ simulation, which is plausible given the enormous variety of evolutions that the planets may have potentially had during a JJ instability phase.

Finally, defining as main belt region the one with $q>$ 1.9 au and $a<$ 3.2 au, we find that $\sim$78\% of the main belt population is lost during the planetary instability.
An additional $\sim$43\% was lost within the first 100 My after the planetary instability due to the inclusion of the terrestrial planets.
Another $\sim$25\% is lost in the subsequent evolution, totalizing $\sim$57$\%$ of material loss after the planetary instability.
In terms of mass, considering that the current MB has about $\sim$5$\times$10$^{-4}$ M$_{\oplus}$ \citep{demeo2013}, our predictions suggest that within the original MB region there should be about $\sim$53$\times$10$^{-4}$ M$_{\oplus}$, about 10 times more mass than now. 
Even so, $\sim$53$\times$10$^{-4}$ M$_{\oplus}$ is very little mass, so that the scenario proposed in this paper is consistent with the Low Mass Asteroid belt hypothesis \citep{levison2015b,ogihara2015,moriaty2015,izidoro2015,drazkowska16,izidoro2016,raymond2017a,raymond2017b}.

We have shown that a JJ instability produces roughly an order of magnitude mass depletion in the MB. \citet{clement2018}'s simulations started from a smooth radial surface density distribution and included about an Earth-mass in the MB.  If we simply applied our depletion factor, \citet{clement2018}'s setup would produce a MB two orders of magnitude more massive than the present-day one.  However, it is worth noting that a second depletion mechanism acts on a more massive belt.  Gravitational self-stirring by resident planetary embryos excites and depletes the belt~\citep{petit01,chambers01b}.  Of course, a more massive primordial belt \citep[as modeled by][]{clement2018} must undergo much stronger depletion than a primordial low-mass belt (as we have modeled). The simulations of \citet{clement2018} did not have high enough resolution to determine the final MB mass; rather, their success criterion was simply that no embryos could survive in the belt~\citep[see Fig. 1 in][]{raymond2009}. It remains to be seen whether a combination of self-stirring and an early JJ instability can provide the requisite depletion.

\section{Conclusion}\label{sec7}

We have shown that a jumping-Jupiter (JJ) evolution during the giant planet instability can excite a dynamically cold primordial main asteroid belt (MB) to an over-excited state (comparable to those obtained in both the Grand Tack model \citep[GT --][]{walsh2011} and in the chaotic excitation \citep{izidoro2016}), which subsequently evolves to the current level of excitation due to the preferential removal of the most dynamically excited asteroids over the Solar System age.  

We started by performing instability simulations of the giant planets starting from an initially 5 planet multi-resonant configuration \citep[with resonant period ratios of 3:2, 3:2, 2:1, 3:2 --][]{deienno2017}, as suggested by hydrodynamical simulations of planet migration in a gas-dominated disk.
We selected a ``nominal'' simulation, which satisfies all constraints already considered for the JJ instability scenario \citep{nesvorny2012,deienno2017}.
Then, we restricted our attention to the JJ instability period and recorded for every 1 year output the orbits of Jupiter, Saturn, and $pl5$ \citep[the ejected extra ice giant planet predicted by][and \citet{nesvorny2012}]{nesvorny2011,batygin2012}. We interpolated the recorded orbits of the planets to simulate their effect over an initially dynamically cold MB.

We found that Jupiter is dynamically responsible for exciting the entire MB. The other planets, Saturn and $pl5$, have only minor direct effects upon the MB, but play an important role in the excitation mechanism by making Jupiter acquire a rapidly varying non-negligible orbital inclination during the instability phase.

The mechanism that excites the MB from a cold initially state to a very excited one is the presence of large and rapidly evolving forced vectors of eccentricity ($e_{forced}$) and inclination ($I_{forced}$) due to the eccentric, inclined and rapidly changing orbit of Jupiter. Because the secular phases of the asteroids are rapidly randomized, different asteroids achieve different amplitudes of oscillation during the JJ instability phase.
Thus, asteroids spread all over the parameter space of orbital eccentricity and inclination. 
Once the instability has ended and Jupiter and Saturn reach their present regular orbits, the ($e_{forced}$) and ($I_{forced}$) forced vectors' reduce in amplitude and evolve regularly. Consequently,  the asteroids are frozen with their acquired proper eccentricities and inclinations.
Our mechanism is similar to that presented by \citet{izidoro2016} but in our case we do not require that Jupiter and Saturn remain in the 2:1 MMR for long time. Our scenario is consistent with Jupiter and Saturn being originally in the 3:2 MMR \citep{masset01}. 

Our results also suggest that the kind of evolution that Jupiter has to have during the JJ instability to excite the MB is more consistent with an early instability in the Solar System rather than a late instability, although new constraints \citep{marty2017} indicate that the instability nevertheless postdated terrestrial planet formation. Finally, we showed that the subsequent evolution of the excited MB throughout the age of the Solar System makes the final distribution of the asteroids quite consistent with the present day asteroid main belt orbital configuration.

We find that if the asteroid belt had originally comprised a large mass, as assumed in \citet{clement2018} from the minimum-mass solar nebula model, the giant planet instability alone would not have removed enough mass from the MB region. In this case, an additional depletion mechanism as the Grand Tack or with some temporary embedded embryos within the MB \citep{raymond2009,clement2018} should be invoked. Still, it remains to be demonstrated whether a combination of \citet{clement2018} model with temporary embedded embryos in the MB and self-stirring from a massive disc with an early JJ instability like shown in this work can provide the required depletion. On the other hand, our results support the Low Mass Asteroid Belt model \citep{izidoro2015,drazkowska16,izidoro2016,raymond2017a,raymond2017b}, which may provide a coherent alternative to the Grand Tack model for the evolution of the inner Solar System.

\acknowledgments

R.D. acknowledges support provided by grant $\#$2014/02013-5, S\~ao Paulo Research Foundation (FAPESP) and CAPES. A. I. thanks financial support provided by grant $\#$2016/12686-2 and $\#$2016/19556-7, S\~ao Paulo Research Foundation (FAPESP) and CAPES.

\software{Mercury \citep{chambers1999}}

\end{document}